\documentstyle[12pt]{article}

\def\beq{\begin{equation}}
\def\eeq{\end{equation}}

\def\noi{\noindent}

\begin{document}
\begin{center}
{\Large\bf Percolation of color sources and critical temperature}
\vspace{0.5 cm}

{\bf  J. Dias de Deus$^1$ and C. Pajares$^{2}$}\par
{\it $^1$CENTRA, Instituto Superior T\'ecnico, 1049-001 Lisboa, Portugal}\par
{\it $^{2}$Departamento de F\'{\i}sica de  Part\'{\i}culas, Universidade de Santiago de Compostela,\\ 
15782--Santiago de Compostela, Spain}\par

\end{center}

\vspace{1 cm}

\noi {\bf Abstract}

We argue that clustering of color sources, leading to the percolation transition,
may be the way to achieve deconfinement in heavy ion collisions. The
critical density for percolation is related to the effective critical
temperature of the thermal bath asociated to the presence of strong
color fields inside the percolating cluster. 
We find that the temperature is rapidity, centrality and energy dependent.
We emphasize the similarities of percolation of strings with color glass condensate.

\newpage
Recently \cite{Ref1}\cite{Ref2} it has been proposed a new
thermalization scenario for heavy ion collisions
which at sufficiently high energies inplies the phase
transition to the quark-gluon plasma. The key ingredient
is the Hawking-Unruh effect \cite{Ref3}\cite{Ref4}. It
is well known that the black holes evaporate by quantum pair
production, and behave as if they have an effective temperature of
\begin{equation}
	T_H=\kappa/2\pi
\end{equation}
where $\kappa=(4 GM)^{-1}$ is the acceleration of gravity
at the surface of a black hole of mass M. The rate of
pair production in the gravitational background of the black
hole can be evaluated by considering the tunneling throuh the 
event horizon. The imaginary part of the action for this clasically
forbidden process corresponds to the exponent of the Boltzmann factor
describing the thermal emision \cite{Ref5}. Unruh showed that a
similar effect arises in a uniformly accelerated frame, where an
observer detects an thermal radiation with the temperature
\begin {equation}
	T_U=a/2\pi
\end {equation}
where a is the acceleration. The event horizon in this case emerges
through the existence of causally disconnected regions of space-time \cite{Ref6}.

Let us now turn to hadronic interactions. The probability to produce
states of masses M due to the chromo-electric fiel E and color
charge $g$ is given by the Schwinger mechanism \cite{Ref7}
\begin {equation}
	W_M\sim exp(-\pi M^2/gE)\sim exp(-M/T)
\end {equation}
which is similar to the Boltzmann weight in a heat bath with
an effective temperature
\begin {equation}
T=\frac{a}{2\pi}\quad,\quad a=2gE/M.
\end {equation}

The full probability P to produce any state of invariant mass M is
\begin {equation}
P(M)=2\pi W_M \rho(M),
\end {equation}
being $\rho$ the density of hadronic final states. In the
dual resonance model, $\rho$ is given by
\begin {equation}
\rho(M)\sim exp\left(M\sqrt{\frac{b}{6}}\right),
\end {equation}
where b is related to the string tension $\sigma$ by the relation
$b=1/2\pi\sigma$. Unitarity implies that the sum of $P(M)$ over
all finite states should be finite, what means
\begin {equation}
T=\frac{a}{2\pi}\leq \frac{1}{4\pi}\sqrt{\frac{6}{b}}\equiv T_{Hag}.
\label{ec7}
\end {equation}

The quantity of the right hand side of (\ref{ec7}) is the well
known Hagedorn temperature, the limiting temperature of hadronic matter
above which hadronic matter under goes a phase transition into the deconfined
phase \cite{Ref8}. So, it also can be considered as the existence
of a limiting acceleration $a_0=\sqrt{3/2b}$ corresponding to
a strong color field. The existence of a limiting
temperature is not exclusive of the dual resonance model, but it
occurs in any system whose exponential mass spectrum has combinatorics
structure \cite{Ref9}.

Heavy ion collisions are currently described in terms of color
strings stretched between the nucleons of the projectile and the target,
which decay into new strings through $q-\bar{q}$ pairs production and 
subsequently hadronize to produce observed hadrons \cite{Ref10}.
Color strings may be viewed as small discs in the transverse space,
$\pi r^2_0 , r_0 \simeq 0.2 fm$, filled with the color field
created by the colliding partons. Particles are produced by the
Schwinger mechanism, emitting $q-\bar{q}$ pairs in this field.
With growing energy and/or atomic number of colliding nuclei,
the number of strings grows and they start to overlap forming
clusters. At a critical density a macros copic cluster
appears that marks the percolation phase transition \cite{Ref11}.
The strong color field inside this large cluster produces an
acceleration which can be seen as a thermal temperature T, by
means of the Hawking-Unruh effect.

In this paper, we compute this temperature to check if its value
is above tre critical phase temperature found in lattice QCD \cite{Ref12}
which is 170-185 MeV depending on the number of flavour (2 or 3) and
for vanishing chemical potentials.

In two dimensional percolation theory the relevant variables is the density
\begin {equation}
	\eta= N_S\ S_1/S_A
\end {equation}
where $N_S$ is the number os strings formed in the collisions and $S_!$
and $S_A$ are the transverse area of one string an the nuclear
overlapping area respectively. Thus, it depends on the impact parameter
$b$ of the collision. For very central collisions $b=0$ and $S_A=\pi R^2_A$.

The critical point for percolating is $\eta_C=1.18-1.5$, depending on the
type of the employed nuclei profile functions, homogeneous or
Wood-Saxon \cite{Ref13}.

We assume that a cluster of N strings behave as a single string
with energy momentum that corresponds to the sum of the
energy-momentum of the overlapping strings and with a stronger color
field resulting from the vectorial sum of the color charges 
$\stackrel{\rightarrow}{Q_{1}}$ of each individual string. The
obtained color field covers the area $S_N$ of the cluster.
As the individual string colors are random oriented in color space,
the average 
$\stackrel{\rightarrow}{Q_{1i}}.\stackrel{\rightarrow}{Q_{1j}}$
is zero, such that if $S_N=S_1$ then $Q^2_N=NQ^2_1$ or
$Q_N=\frac{1}{\sqrt{N}}NQ_1$
with a color reduction factor $1/\sqrt{N}$. 
In general, $S_N \geq S_1$ and the color reduction becomes
$\sqrt{\frac{1}{N}\frac{S_N}{S_1}}$ \cite{Ref14}.

In the framework of the Schwinger model extended to color fields, 
the particle density $\mu_N$ is proportional to the
color charge $Q_N$ and the $<p^2_T>_N$ is proportional to
the string tension $\sigma_N, \mu_N \sim Q_N$, and
$<p^2_T>_N \sim \sigma_N$. On the other hand, from Gauss theorem

\begin {equation}
Q_N \sim S_N \sigma_N
\end {equation}
we thus obtain
\begin {equation}
\mu_N=\sqrt{\frac{1}{N}\frac{S_N}{S_1}}N\mu_1 \quad and \quad
<p^2_T>_N=\sqrt{N\frac{S_1}{S_N}}<p^2_T>_1
\label{ec10}
\end {equation}

The ratio of the two equations (\ref{ec10}) gives rise to the scaling
law
\begin {equation}
	<p^2_T> S_N/\mu_N = <p^2_T>_1 S_1/\mu_1
\end {equation}
valid for all projectiles, targets and energies. This scaling law
is in reasonable agreement with data \cite{Ref15}. A similar scaling
is found in the Color Glass Condensate (CGC) framework \cite{Ref16}.

Asymtotically, for a random distribution of $N_S$ strings,

\begin {equation}
	<\frac{1}{N}\frac{S_N}{S_1}> \equiv F^2(\eta)
\end {equation}
with
\begin {equation}
F(\eta)=\sqrt{\frac{1-e^{-\eta}}{\eta}}
\end {equation}
being the color reduction factor. It follows that
\begin {equation}
\mu=F(\eta)N_S \mu_1 \quad and \quad <p^2_T>=<p^2_T>_1/F(\eta)
\label{ec14}
\end {equation}
Note that $\sigma_N=\sigma_1/F(\eta)$.

The tension of the macroscopic cluster fluctuates around its
mean value because the chromoelectric field is not strictly constant.
Assuming a gaussian form for these fluctuations \cite{Ref17}
we have for the transverse momentum distributions
\begin {equation}
\frac{dn}{d^2 p_\bot}\sim \sqrt{\frac{2}{<x^2>}}
\int^{\infty}_{0}dx\,exp(-x^2/2<x^2>) exp\left(-\pi \frac{p^2_\bot}{x^2}\right)
\end {equation}
which gives rise to the thermal distribution
\begin {equation}
\frac{dn}{d^2 p_\bot}\sim exp\left(-p_\bot \sqrt{\frac{2\pi}{<x^2>}}\right)
\end {equation}
with
\begin {equation}
<x^2>=\pi <p^2_T>_1/F(\eta)
\end {equation}

Therefore,
\begin {equation}
T=\sqrt{<p^2_T>_1}/\sqrt{2F(\eta)}
\end {equation}
If we identifiy the percolation transition temperature to the Hagedorn
temperature, we obtain
\begin {equation}
T_H=\sqrt{\frac{<p^2_T>_1}{2F(\eta_C)}}
\end {equation}
which gives for $\sqrt{<p^2_T>_1}\simeq 250 MeV$, and with
$\eta \simeq 1.18-1.5,\ T_C\simeq 200-250 MeV$.

Similar value has been obtained previously in the framework
of the CGC \cite{Ref1},\cite{Ref18}. In this case
$T=Q_S/2\pi$ where $Q_S$ is the saturation momentum.
As it has been pointed out, the role of the scale $Q_S$ in
the transverse momentum distributions is played by 
$<p_T>_1/F(\eta)$ in percolation. 
Similar phenomenology can be obtained in both schemes \cite{Ref19}.

Most of the considerations presented in the framework of the
CGC \cite{Ref18} related to a rapid thermolization can be
applied to our approach. Hydrodynamics models \cite{Ref20}
have sucessfully described the collective flow of the produced
hadrons and the low $p_T$ single particle spectra. 
This sucess has led to the conclusion that the created matter
behaves like an ideal liquid with almost negligible
viscosity \cite{Ref20}\cite{Ref21}. The existing
microscopic approaches are not able to explain a fast
thermalization. For instance, a transport model based on
independent strings \cite{Ref22} does not present early
thermalization although the corresponding energy density
profiles at proper time $\tau=1 fm/c$ compare well with
hydrodynamics assumptions for initial energy density distributions.
At this point, the interaction among string, forming a large
cluster with strong color field inside helps to reach a fast
thermalization.

In fact, below the percolation threshold we have clusters
with a few number of strings and different occupied areas what
give rise to different temperatures. Above the threshold,
due to the high density, there are strong interactions among the
strings forming essentially only one large cluster. The
fragmentation of this cluster give rise to a thermal distribution
of the produced particles. The interaction among these large
number of strings can be considered as an strong interaction
at partonic level which give rise to a color composition and
the formation of one cluster.

Similarly to what happens in the CGC framework, we can ask
for the charasteristic time over which the chromo-electric
field changes. In our case, it would be proportional to the
momentum scale, i.e. $<p_T>_1/F(\eta)$, which implies a lifetime
of the cluster, $\tau=1/T \simeq 0.7-1 fm$ for 
$\eta \geq \eta_C$. As the string density increases $\tau$ decreases.
Similar conclusions can be obtained, in a more detailed model
as it is discussed in ref \cite{Ref23}.

In the CGC the temperature depends on the rapidity provides that
$Q_S$ depends on $y$. This dependence may trigger instablility
of the system and speed up thermalization process. In percolation,
the temperature (average transverse momentum) depends also on the
center of mass rapidity $y$. In fact, the first equation in (\ref{ec14}),
making $\mu \rightarrow \frac{dn}{dy}$ and $\eta_1 \rightarrow \bar{n}$,
can be rewritten in the form
\begin {equation}
	\frac{dn}{dy}=F(\eta)\overline{N}_S(\Delta, y_b)\bar{n}
	\label{ec20}
\end {equation}
where $\Delta=y-y_b , \frac{\sqrt{s}}{2} \sim exp(y_b)\ 
\textrm{and}\  \eta \equiv \left(\frac{r_0}{R}\right)^2 \bar{N_S}(\Delta,y_b)$.
Note that, in general, $\bar{N_S}$ is a function of $y$ and $y_b$
while $\bar{n}$ is not.

For $\Delta \simeq 0$, the rapidity distribution reflects the 
scaling behaviour of the parton structure functions of the
valence strings, their number being proportional to the 
number of nucleon participants, $N_{part}=2N_A$:
\begin {equation}
	\Delta \simeq 0 \quad
	\overline{N}_S(0, y_b)\rightarrow 
	\overline{N}_S(0)\simeq N^P_S (0) N_A
	\label{ec21}
\end {equation}
where $\bar{N}^P_S$ is the number of strings in $pp$ collisions.

For $\Delta \simeq -y_b$, we are in the domain of sea
partons and sea string contribution, their number being proportional
to the number of nucleon-nucleon collisions, $N_A^{4/3}$:
\begin {equation}
	\Delta \simeq -y_b \quad
	\overline{N}_S(-y_b, y_b)\rightarrow 
	\overline{N}_S(y_b)\simeq \overline{N}^P_S (y_b) N_A^{4/3}
	\label{ec22}
\end {equation}
>From equations (\ref{ec20}),(\ref{ec21})and (\ref{ec22}) it follows
\begin {equation}
\Delta \simeq 0: \frac{dn}{dy}\sim N_A \frac{1}{N_A^{1/6}}\ 
 \textrm {independent of}\ y_b
\end {equation}
\begin {equation}
\Delta\simeq -y_b: \frac{dn}{dy}\sim N_A,\  \textrm {increasing with energy}
\end {equation}
Regarding the $p_T$ behaviour we have
\begin {equation}
\Delta \simeq 0: <p_T>\simeq \eta^{1/4}\simeq N_A^{1/12},\  \textrm {independent of}\ y_b
\end {equation}
\begin {equation}
\Delta\simeq -y_b: , <p_T>\simeq \eta^{1/4}\simeq N_A^{1/6}\  \textrm {increasing with}\ y_b
\end {equation}

In this way, the temperature is larger in central rapidity than in
the extremes in a factor $N_A^{1/12}$, which for central $Au-Au$
collisions is around $1.6$. This difference increases with
energy. This considerable difference of temperature in different
rapidity slices generater a viscosity, which is very small.
Following \cite{Ref18}, the shear viscosity can be estimated by
\begin {equation}
\frac{\eta_S}{n}=<p_T>\lambda \sim <p_T>\frac{1}{nS_1}=
\frac{<p_T>_1 L}{\eta F(\eta)}
\end {equation}

where we use for $n$, the number of strings per unit of volume,
$n=\frac{N_S}{\pi R^2_A L}$ being L the longitudinal extension,
$L\simeq 1 fm$. 
For $\eta \geq \eta_C$, we obtain a value around 1 decreasing
with the density.

We thank N. Armesto, M. Braun, E.G. Ferreiro, E. Levin and
H. Satz for discussion. This work has been partially done
under contracts POCI/FIS/55759/2004 (Portugal) and
FPA 2005-01963 (Spain) and PGIDIT03PX1C20612PN (Galicia).

\end{document}